\documentclass[twocolumn,aps,showpacs,prb,superscriptaddress]{revtex4-2}

\usepackage{hyperref}
\usepackage{natbib}
\usepackage{graphicx}
\graphicspath{ {./images/} }
\usepackage{physics}
\usepackage[utf8]{inputenc}
\usepackage[english]{babel}
\usepackage{xcolor}
\usepackage[normalem]{ulem}

\begin{document}

\title{Renormalization group study of systems with quadratic band touching}

\author{Jeet Shah}
\affiliation{Department of Physics, Indian Institute of Science, Bangalore 560 012, India}
\author{Subroto Mukerjee}
\affiliation{Department of Physics, Indian Institute of Science, Bangalore 560 012, India}


\begin{abstract}
Lifshitz transitions in two 2D systems with a single quadratic band touching point as the chemical potential is varied have been studied here. The effects of interactions have been studied using the renormalization group (RG) and it is found that at the transition a repulsive interaction is marginally relevant and an attractive interaction is marginally irrelevant. We corroborate the results obtained from the RG calculation by studying a microscopic model whose ground state and Green's functions can be obtained exactly. We find that away from the transition, the system displays an instability towards forming and excitonic condensate. 
\end{abstract}

\pacs{72.10.-w, 71.45.-d}

\maketitle

\section{Introduction}

A quadratic band touching point is a point in the Brillouin zone where two bands touch each other quadratically. Such bands are obtained in Bernal stacked bilayer graphene within the nearest neighbor tight binding approximation \cite{mccann2013blg, nilsson2006interactions, partoens2006estructure, nandakishore2010qahblg, nandkishore2012blg,koshino2017mlg}. The two low energy bands touch each other quadratically at the two valleys in the Brillouin zone of bilayer graphene. This system is known to show instabilities towards a nematic state driven by the Coulomb interactions \cite{vafek2010blg}. Another system with quadratic band touching is twisted bilayer graphene, which consists of two layers of monolayer graphene twisted relatively to each other. For a generic twist angle, the low energy bands can be approximated by a linear dispersion similar to massless Dirac fermions of monolayer graphene. However, as the angle is varied, the Dirac velocity changes and for a set of discrete angles, called the `magic' angles, the Dirac velocity vanishes completely \cite{bistritzer2011moire, cao2018superconductivity}. At these angles, the bands no longer have a linear dispersion. Instead there are particle-hole symmetric bands which touch each other quadratically \cite{kang2018tblg,hejazi2019tblg, kwan2020tblg}. Twisted bilayer graphene, especially at magic angles has been studied extensively in the last three years after the discovery of unconventional superconductivity and Mott insulating behavior at magic angles \cite{cao2018superconductivity,macdonald2020tblg, sharma2020tblg,stepanov2020nature}.

While the above phenomena occur at finite carrier density, (i.e. the Fermi level is within one of the two touching bands and not at the touching point), interesting phenomena can occur as the carrier density is lowered. In particular, the topology of the Fermi surface can change from that of a closed curve (assuming the system is two dimensional) when the Fermi level is inside a band to that of a point, when it is at the touching point of the two bands. Such transitions where the topology of the Fermi surface changes, upon changing a parameter are generally categorized as Lifshitz transitions \cite{lifhitz1960transition,blanter1994ett} and have been known to be accompanied by interesting physical effects such as anomalous transport~\cite{chi2017transport,nandy2016transport}, anomalous thermodynamic properties~\cite{hackl2011lt, dagens1979lt}, effective mass enhancement~\cite{fink2016mass}, and reenterant superconductivity~\cite{sherkunov2018fesc}. Thus, it is interesting to investigate the nature of this particular type of Lifshitz transition and concurrent physical effects.

It should be noted the above change of topology of the Fermi surface is the same regardless of whether the band touching is quadratic or linear (as for a Dirac dispersion). However, the features of the transition can be quite different owing to kinematic considerations. Specifically, in the case of linear band touching, the density of states at the touching point goes to zero as the touching point is approached \cite{castro2009graphene}. As a result, the Coulomb interactions are not screened and a QED-like field theoretic treatment is required to account for the long-range interactions \cite{vozmediano2011rg,gonzlez1994graphenerg}. On the other hand, when the bands touch quadratically, screening simplifies the form of the interaction and thus also the form of the field theory required to understand the transition. This is one of the main motivations for us to consider only the latter situation in this work. Physical systems with quadratic band touchings have been studied earlier in the Refs.~\cite{giovannetti2015qbc,chong2008quaddegen, fu2011tci, huang2016qb}. We note that it is also possible to engineer physically realistic models arbitrary power law dispersion at the band touching point in certain physically interesting types of lattice models \cite{haldar2018SYK}.

Two continuum models in two dimensions with quadratic band touching are studied in this paper. The first resembles one of the valleys of bilayer graphene and the second can be obtained in the low energy limit of an exactly solvable microscopic model. As discussed above, the Coulomb interactions are short ranged and can be accounted for with the Renormalization Group (RG) formalism by a single coupling constant. The flow equation of this coupling constant is obtained. It is found that repulsive interactions are marginally relevant, while attractive interactions are marginally irrelevant in the RG sense. This is in stark contrast to what happens when Fermi energy is away from the band touching point, where the Fermi wave-vector is finite and the Fermi surface is a circle. It is know that such a system is a Fermi liquid \cite{shankar1994rg}, with repulsive interactions being marginally irrelevant and attractive interactions being marginally relevant. Moreover, the attractive interactions, if present, drive the system away form the Gaussian fixed point towards a superconducting state.
This difference in behaviour depending on the topology of the Fermi surface is the signature that we find of this particular type of Lifshitz transition. The Green's functions are calculated at one-loop level for the two models studied. 

Microscopic models with quadratic touching protected by symmetry have been studied at the mean field level in the Refs. \cite{fradkin2009quadratic, uebelacker2011qbcp, pawlak2015qbcp, murray2014qbcp, dora2014flatparabolicband, liu2010kagome, wen2010kagome, tsai2015lieb}. The microscopic model studied here while not affording any topological protection to the band touching has the advantage of simplicity over these other models, which makes it possible to obtain exact results which can be compared to those obtained from the field theory. In particular, we have determined the exact ground state and the exact Green's  function in the presence of repulsive interactions.

The microscopic model studied also exhibits a Lifshitz transition as a microscopic parameter (a Zeeman field), is varied as opposed to the Fermi energy. While the properties of the transition remain the same as in the case where the Fermi energy is varied, away from the transition, we find that on one side, the system possesses two types of instabilities, a BCS instability and excitonic condensate instability \cite{keldysh1968exciton} and on the other side, it is an insulator.

The rest of the paper is organized as follows: in Sec. \ref{sec:hamiltonians}, we give the Hamiltonians and the imaginary time action for the two models studied in this paper. In Sec. \ref{sec:rg_analysis}, we analyze these models using RG and determine the RG flow equation of the coupling constant followed by calculation of the Green's functions for the two models.  In Sec. \ref{sec:microscopic_model}, we present an exactly solvable microscopic model which has two bands touching quadratically. We determine the ground state and the Green's function in the presence of interactions. We also comment on the excitations of this microscopic model. In Sec. \ref{sec:another_lt}, we analyze the Lifshitz transition as the Zeeman field is varied. We derive the RG flow equations for intraband and interband couplings, and study the excitonic condensate instability. Finally, in Sec. \ref{sec:discussion} we discuss our results.

\section{The Hamiltonians} \label{sec:hamiltonians}

\begin{figure}
	\centering
	\includegraphics[width=0.95\columnwidth]{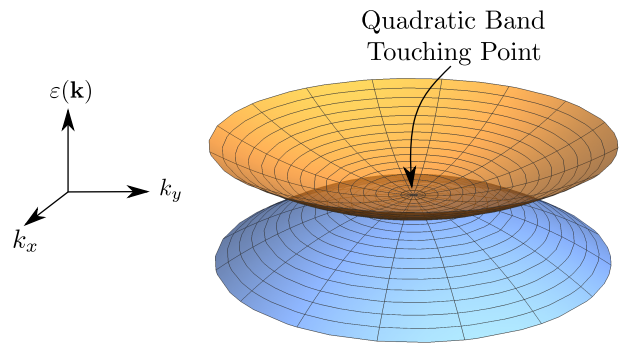}
	\caption{Two bands with quadratic dispersions and the quadratic band touching point where they touch each other.}
	\label{fig:qbcp}
\end{figure}

We start by introducing the two low energy models that will be analyzed using the renormalization group later. These are models in two dimensions which have a single point where two bands touch each other quadratically and isotropically. The curvatures of the two bands are equal in magnitude and opposite in sign. The first model resembles the low energy Hamiltonian of one of the valleys of bilayer graphene. This model will be referred to as model A and has the following Hamiltonian in the small $|\textbf{k}|$ limit.
 \begin{equation} \label{eq:2_band_xy}
H_A(\textbf{k}) = \begin{pmatrix}
0 & \frac{k^2}{2m}e^{-2i\theta} \\
\frac{k^2}{2m}e^{2i\theta} & 0
\end{pmatrix}
\end{equation}
where, $\theta = \tan^{-1}(k_y/k_x)$ and $k = \sqrt{k_x^2+k_y^2}$. There are two bands with dispersions $\varepsilon_{\pm} (\textbf{k}) = \pm \frac{k^2}{2m}$ as shown in Fig. \ref{fig:qbcp}. 

The other model that we have studied is described by a Hamiltonian with off-diagonal matrix elements equal to 0. It is interesting to study this model because it is obtained in the continuum limit of an exactly solvable microscopic model as we discuss later in Sec. \ref{sec:microscopic_model}. This model will be referred to as model B and has the following Hamiltonian.
\begin{equation} \label{eq:2_band_z}
H_B(\textbf{k}) = 
\begin{pmatrix}
\frac{k^2}{2m} & 0 \\
0 &  -\frac{k^2}{2m}
\end{pmatrix}
\end{equation}
Clearly, this model too has bands with quadratic dispersions identical to model A.

The imaginary time action plays a central role in the renormalization group transformations. Consider the situation where the chemical potential is at the quadratic band touching point, so that the Fermi surface is a single point. Since, we are concerned with the low energy or long wavelength properties of the system, we keep only the modes with momenta within a distance $\Lambda$ about the Fermi point.  If the size of system is $L$, then the cutoff wavelength $\Lambda$ is of the order $1/L$. This is equivalent to stating that $\Lambda$ is much smaller than the size of the Brillouin zone.  The modes under consideration form a disk of radius $\Lambda$ in the Brillouin zone. This is in contrast with a Fermi liquid where the modes under consideration form an annular region of width $\Lambda$ on both sides of the Fermi circle (or surface in 3D) \cite{shankar1994rg}.

The non-interacting or the quadratic part of the action is given by
\begin{equation}
S_0 =  \int_{|\textbf{k}|<\Lambda} \frac{d\textbf{k}}{(2\pi)^2} \int_{-\infty}^\infty \frac{d\omega}{2\pi}   \psi^\dagger(\textbf{k},\omega) \left( -i\omega + H(\textbf{k}) \right) \psi(\textbf{k}, \omega)
\nonumber
\end{equation}
where $H(\textbf{k})$ could either be the Hamiltonian of model A or B, $\psi(\textbf{k},\omega)$ is a Grassmann field with two components and $\psi^\dagger(\textbf{k},\omega)$ is its conjugate.
The quadratic dispersion implies that the density of states of each band at $\varepsilon=0$ is finite. The non-zero density of states at the Fermi energy implies that the Thomas-Fermi wave-vector is finite and thus, the long ranged Coulomb interaction are screened and hence are short ranged. This affords a simplification of treatment while still allowing us to investigate the physics of the particular type of Lifshitz transition we are interested in. With long ranged Coulomb interactions such as in systems like monolayer graphene with linearly touching bands, a QED like treatment can be employed with a dynamical gauge field in addition to fermionic Grassmann fields to describe the electrons \cite{vozmediano2011rg,gonzlez1994graphenerg}.

The quadratic dispersion of the Hamiltonian implies that the Grassmann fields have a scaling dimension of $-3$ at the Gaussian fixed point. This implies that in a Taylor expansion in ${\bf k}$ and $\omega$ of the quartic interaction terms, all terms aside from the first term are irrelevant at the fixed point and the first term is marginal.  Thus, unlike in a Fermi liquid there are only coupling \textit{constants} and not coupling functions. The antisymmetry of the Grassmann fields restricts the number of coupling constants to just one. It is important to realize that this reduction of the coupling function to a few coupling constants is made possible because the Fermi surface is a point.
The interaction part of the action can thus be written as
\begin{equation} \label{eq:S_I}
S_I = \frac{g}{2}  \int_{\textbf{k},\omega} \bar{\psi}_1(4) \bar{\psi}_2(3) \psi_1(2) \psi_2(1)
\end{equation}
where $g$ is the coupling constant, the numbers 1 through 4 in the argument of the fields represent pairs of momenta and frequencies. For example, $\psi_1(2)$ stands for $\psi_1(\textbf{k}_2,\omega_2)$. The following shorthand is used
\begin{multline} \label{eq:measure}
\int_{\textbf{k}, \omega} = \int_{|\textbf{k}_i| < \Lambda} \prod_{i=1}^4\left[\frac{d \textbf{k}_i d \omega_i}{(2\pi)^3} \right] (2\pi)^3\delta(\textbf{k}_4 + \textbf{k}_3 - \textbf{k}_2 - \textbf{k}_1) \\ \delta(\omega_4 + \omega_3-\omega_2-\omega_1) 
\nonumber
\end{multline}
The notation used in this paper is similar, although not identical to the notation of Ref. \cite{shankar1994rg}. $S_I$ represents the density-density interaction of strength $-g/2$. The total action is the sum of the quadratic and the quartic pieces. 
\begin{equation}\label{eq:general_S_I} 
S = S_0 + S_I 
\nonumber
\end{equation}

\section{Renormalization Group analysis} \label{sec:rg_analysis}

\subsection{One-loop analysis}

Following the steps outlined in Ref. \cite{shankar1994rg}, the fields are divided into two parts, slow modes denoted by $\psi_s$ and fast modes denoted by $\psi_f$. The slow modes have momentum less than $\Lambda/s$, while the fast modes have momentum in the range $\Lambda/s<k<\Lambda$. Integrating out the fast modes changes the effective action of the slow modes and the coupling constant gets renormalized. The partition function as a path integral is given by
\begin{equation}
Z = \int  \left[ \mathcal{D}\bar{\psi} \mathcal{D}\psi \right] e^{- S_0 - S_I}
\nonumber
\end{equation}
Next, $S_I$ is treated as a perturbation to $S_0$.
\begin{align} \label{eq:partition_fun}
Z & =  \int  \left[ \mathcal{D}\bar{\psi_s} \mathcal{D}\psi_s \right] \left[ \mathcal{D}\bar{\psi_f} \mathcal{D}\psi_f \right] e^{- S_0(\psi_s)} e^{- S_0(\psi_f)}  e^{ - S_I(\psi_s, \psi_f) } \nonumber \\
& = \int  \left[ \mathcal{D}\bar{\psi_s} \mathcal{D}\psi_s \right] e^{- S_0(\psi_s)} \left< e^{ - S_I(\psi_s, \psi_f) } \right>_0 \nonumber\\
& \approx  \int  \left[ \mathcal{D}\bar{\psi_s} \mathcal{D}\psi_s \right] e^{- S_0(\psi_s)} e^{ -\left< S_I \right>_0 + \frac{1}{2} \left[ \left< S_I^2 \right>_0 - \left< S_I \right>_0^2 \right]}
\nonumber
\end{align}
Here $\langle \cdots \rangle_0$ denotes the expectation value with respect to the fast modes. To calculate the flow of the coupling $g$ and any other terms which might be generated from the RG transformations, we need to compute 
\begin{equation}\label{eq:delta_S}
\delta S = \left< S_I \right>_0 - \frac{1}{2} \left[ \left< S_I^2 \right>_0 - \left< S_I \right>_0^2 \right]
\end{equation}
It is useful to have the expressions of the non-interacting Green's function for models A and B which are needed to evaluate the expectation values using Wick's theorem. The non-interacting Green's function is given by
\begin{equation}
G(\textbf{k} \omega) = [ -i\omega +H(\textbf{k}) ]^{-1}
\nonumber
\end{equation}
For model A this evaluates to
\begin{equation} \label{eq:gf0_a}
G(\textbf{k} \omega) = \frac{1}{\omega^2+\left(\frac{k^2}{2m}\right)^2}
\begin{pmatrix}
i\omega & k^2e^{-2i\theta}/2m\\
k^2e^{2i\theta}/2m & i\omega
\end{pmatrix}
\end{equation}
and for model B, it evaluates to
\begin{equation} \label{eq:gf0_b}
G(\textbf{k}\omega) = 
\begin{pmatrix}
\frac{1}{-i\omega+k^2/2m} & 0\\
0 & \frac{1}{-i\omega -k^2/2m}
\end{pmatrix}
\end{equation}

\begin{figure}
	\centering
	\includegraphics[width=\columnwidth]{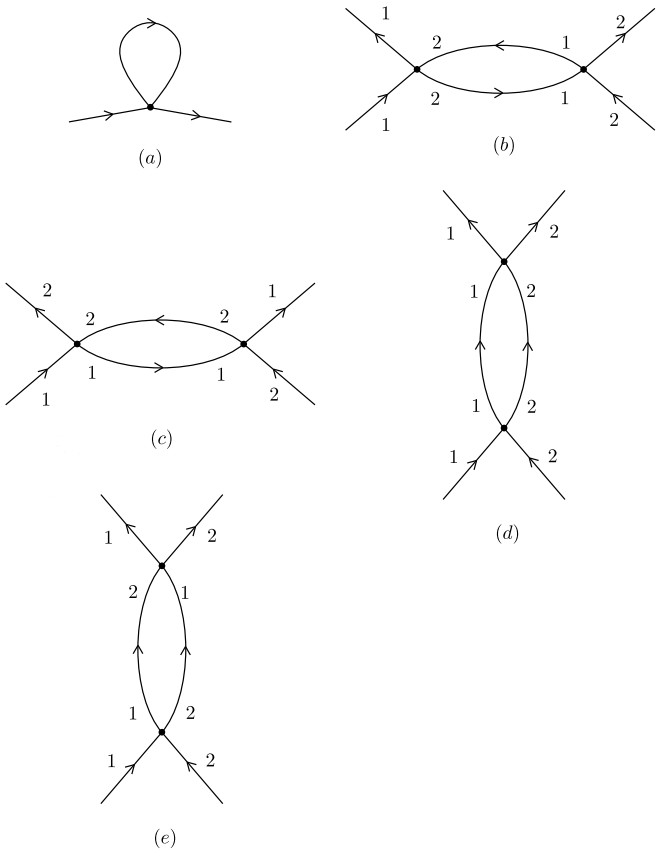}
	\caption{Feynman diagrams. Numbers $1$ and $2$ denote the fermionic species. (a) Tadpole diagram (b) ZS diagram (c) ZS$'$ diagram  (d) and (e) are BCS diagram}
	\label{fig:diagrams}
\end{figure}

The first term in Eq. \ref{eq:delta_S} corresponds to the tadpole diagram shown in Fig. \ref{fig:diagrams} (a). The change in action caused by this can be shown to be 
\begin{widetext}
\begin{equation} \label{eq:tadpole}
\delta S_{T}= \frac{gs^2}{2} \int_{|\textbf{k}|<\Lambda} \frac{d\textbf{k} d\omega}{(2\pi)^3} \psi^\dagger (\textbf{k}\omega)
\int_{f} \frac{d\textbf{k}_1 d\omega_1}{(2\pi)^3}
\begin{pmatrix}
G_{22} (\textbf{k}_1\omega_1) & -G_{12} (\textbf{k}_1\omega_1)\\
-G_{21} (\textbf{k}_1\omega_1) & G_{11} (\textbf{k}_1\omega_1)\\
\end{pmatrix}
\psi(\textbf{k}\omega)
\end{equation}
\end{widetext}
A straightforward substitution and integration gives the contribution of the tadpole diagram for model A as
\begin{align}
\delta S_T= \frac{-g\Lambda^2(s^2-1)}{16\pi} \int_{|\textbf{k}| < \Lambda} \frac{d\textbf{k}d\omega}{(2\pi)^3}
\psi^\dagger (\textbf{k}\omega)
\begin{pmatrix}
1 & 0\\
0 & 1
\end{pmatrix}
\psi(\textbf{k}\omega)
\end{align}
This term represents a change in the chemical potential caused by the interactions. Since our goal is to understand the properties at half filling where the chemical potential is equal to 0, a counter term which cancels the above change in the chemical potential is needed in the starting action. Once we have added such a counter term, we can ignore the above change.

For model B, the contribution of the tadpole diagram is given by
\begin{equation} 
\delta S_T = \frac{-g\Lambda^2(s^2-1)}{8\pi} \int_{|\textbf{k}|<\Lambda} \frac{d\textbf{k} d\omega}{(2\pi)^3} \psi^\dagger (\textbf{k}\omega)
\begin{pmatrix}
1 & 0 \\
0 & 0
\end{pmatrix}
\psi(\textbf{k}\omega)
\end{equation}
This represents a shift in the dispersion of one of the fermionic species by a constant amount.

The term in Eq. (\ref{eq:delta_S}) within the square brackets determines the flow of the coupling constant. Similar to Ref. \cite{shankar1994rg}, there are three types of diagrams which cause a flow of $g$: the ZS, ZS$'$ and the BCS diagrams which are shown in Fig. \ref{fig:diagrams}(b), \ref{fig:diagrams}(c), \ref{fig:diagrams}(d) and \ref{fig:diagrams}(e). The flow of $g$ in these three channels is found  to be
\begin{equation}\label{eq:flow_zs}
\left( \frac{dg}{d\ln s}  \right)_{ZS} = -\frac{g^2}{2} \int_{|\textbf{k}|<\Lambda,\omega} G_{21} (\textbf{k}\omega) G_{12} (\textbf{k}\omega)
\end{equation}
\begin{equation}\label{eq:flow_zs'}
\left( \frac{dg}{d\ln s} \right)_{ZS'} = \frac{g^2}{2} \int_{|\textbf{k}|<\Lambda,\omega} G_{11} (\textbf{k}\omega) G_{22} (\textbf{k}\omega)
\end{equation}
\begin{multline}\label{eq:flow_bcs}
\left( \frac{dg}{d\ln s} \right)_{BCS} = \frac{g^2}{2} \int_{|\textbf{k}|<\Lambda,\omega} [G_{11} (\textbf{k}\omega) G_{22} (-\textbf{k}-\omega) \\
 -G_{21} (\textbf{k}\omega) G_{12} (-\textbf{k}-\omega)]
\end{multline}

From Eqs. (\ref{eq:flow_zs}) $-$ (\ref{eq:flow_bcs}) and the expressions of the Green's function from Eq. (\ref{eq:gf0_a}) and Eq. (\ref{eq:gf0_b}), we find the individual contributions of the ZS, ZS$'$ and the BCS channels to the flow of $g$ in model A as $-g^2 m /(8\pi)$, $-g^2m /(8\pi)$ and 0 respectively. On the other hand, for model B, these contributions are 0, $-g^2m/(4\pi)$ and 0 respectively. The sum of all these channels is the same for both the models giving the total flow equation as
\begin{equation} \label{eq:g_flow_specific2}
\frac{dg}{d\ln s} = -\frac{g^2m}{4\pi}
\end{equation}

\begin{figure}
	\centering
	\includegraphics[width=0.9\columnwidth]{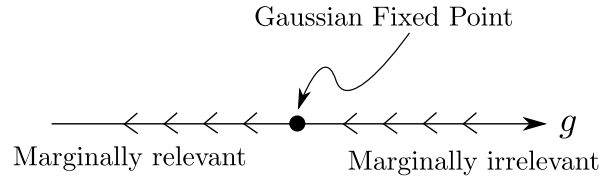}
	\caption{RG flow  of the coupling constant $g$. Positive $g$ is marginally irrelevant and negative $g$ is marginally relevant.}
	\label{fig:flow_g}
\end{figure}

Since the right hand side is always negative, $g$ always decreases as the fast modes are integrated out. This implies that a positive $g$, which corresponds to an attractive interaction decreases. On the other hand, a negative $g$, which corresponds to repulsive interactions increases in magnitude. Thus, we conclude that an attractive interaction is marginally irrelevant and a repulsive interaction is marginally relevant. The flow of the coupling is shown in Fig. \ref{fig:flow_g}. The Gaussian fixed point thus appears to be stable to attractive interactions and unstable to repulsive interactions. This behaviour is completely opposite to a Fermi liquid. We discuss this seemingly peculiar behavior of the coupling in more detail in Sec.~\ref{sec:discussion}. 

\subsection{Green's function}

Starting from the action, the interacting Green's function can be calculated by treating the interaction as a perturbation. The interacting Green's function up to one-loop level is given by
\begin{multline}\label{eq:gf_series}
G_{I,\alpha \beta} (2;1)  \approx  -\langle  \bar{\psi}_\beta(2) \psi_\alpha(1) \rangle_0 +\langle  S_I \bar{\psi}_\beta(2) \psi_\alpha(1) \rangle_0  \\ - \langle S_I \rangle_0 \langle  \bar{\psi}_\beta(2) \psi_\alpha(1) \rangle_0 
\end{multline}
where $\alpha$ and $\beta$ denote the component of the field and take values 1 or 2. Calculating the expectation values using the Wick's theorem gives the following Green's function for model A.
\begin{multline}
G_I(2;1) = \frac{\delta(2-1)}{\omega^2 + (k^2/2m)^2} \times \\
\begin{pmatrix}
i\omega - \tilde{g}\frac{\omega^2-(k^2/2m)^2}{\omega^2+(k^2/2m)^2} & \frac{k^2e^{-2i\theta}}{2m}\left( 1+\frac{2i\omega \tilde{g}}{\omega^2+(k^2/2m)^2} \right) \\
\frac{k^2e^{2i\theta}}{2m}\left( 1+\frac{2i\omega \tilde{g}}{\omega^2+(k^2/2m)^2} \right) & i\omega - \tilde{g}\frac{\omega^2-(k^2/2m)^2}{\omega^2+(k^2/2m)^2}
\end{pmatrix}
\nonumber
\end{multline}
Here, $\tilde{g} = g\Lambda^2/16\pi$. Taking the inverse of the matrix and keeping only $\mathcal{O}(g)$ terms gives
\begin{equation}
G_I^{-1} (2;1) = \delta(2-1) \begin{pmatrix}
-i\omega - \tilde{g} & k^2e^{-2i\theta} /2m\\
k^2e^{2i\theta}/2m & -i\omega -\tilde{g}
\end{pmatrix}
\end{equation}
The diagonal terms represent a chemical  potential of $\tilde{g}$. Therefore, the effect of the interaction at one loop is to change the chemical potential by an amount $\tilde{g}$.

Applying Eq. (\ref{eq:gf_series}) to model B gives the interacting Green's function up to $\mathcal{O}(g)$ as
\begin{equation} \label{eq:gf_field_theory}
G_I(2;1) = \delta(2-1) \begin{pmatrix}
\frac{1}{-i\omega+\frac{k^2}{2m}-\frac{g\Lambda^2}{8\pi}} & 0\\
0 & \frac{1}{-i\omega - \frac{k^2}{2m}}
\end{pmatrix}
\end{equation}
The Green's function calculated above from the field theory will be compared with that from the microscopic model considered in the next section.

\section{Microscopic Model} \label{sec:microscopic_model}

\begin{figure}
	\centering
	\includegraphics[width=\columnwidth]{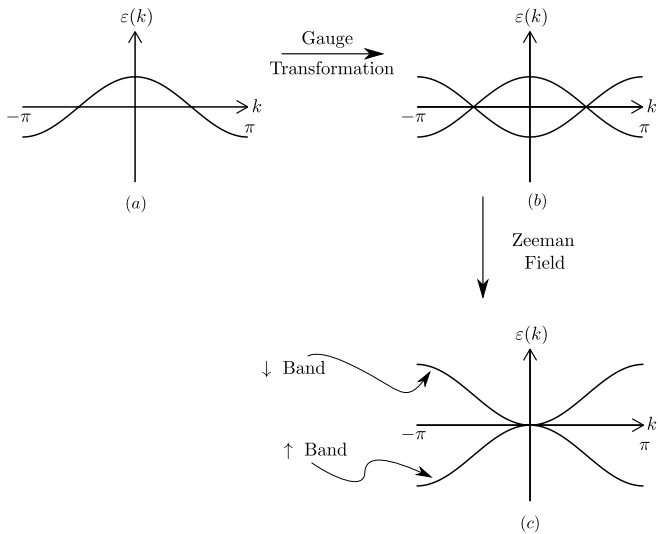}
	\caption{(a) shows two overlapping bands for $\uparrow$ and $\downarrow$ particles. The gauge transformation of Eq. (\ref{eq:gauge_transformation}) changes the sign of the hopping, leading to (b) which shows $\uparrow$ band with hopping $t$ and $\downarrow$ band with hopping $-t$. (c) shows quadratically touching band.}
	\label{fig:band_structure_steps}
\end{figure}
One of the main reasons for studying model B is that it can be realized as the low energy theory of an exactly solvable microscopic model. This model is a system of spin-1/2 fermions hopping on a square lattice with on site interactions of strength $U$ in the presence of a Zeeman field. As will be shown below, the Zeeman field has to be of a particular strength for the two bands of up-spin and down-spin fermions to touch each other quadratically at exactly one point. 

First, we understand how quadratically touching bands can be obtained by a step-wise process. A tight binding model of spin-1/2 fermions with nearest neighbor hopping $t$ has the Hamiltonian $\sum_{\langle i,j \rangle} \left( t c_{i\uparrow}^\dagger c_{j\uparrow} +t c_{i\downarrow}^\dagger c_{j\downarrow} + \text{h.c.} \right)$ and the dispersion  $\varepsilon_{\sigma} (\textbf{k}) = 2t ( \cos (k_x) + \cos (k_y))$, where $\sigma = \uparrow, \downarrow$ denotes the spin. A 1D depiction of these bands is schematically shown in the Fig. \ref{fig:band_structure_steps}(a). The top of the bands can be approximated by a parabola. The two bands have the same curvature, however we want them to have opposite curvatures. One way to achieve this is to flip the sign of hopping of $\downarrow$ spin fermions by performing the following gauge transformation.
\begin{align}
\begin{split}
c_{i\downarrow} \rightarrow c_{i\downarrow} \;\;\; \text{if} \;\;\; i \in A\\
c_{i\downarrow} \rightarrow - c_{i\downarrow} \;\;\; \text{if} \;\;\; i \in B \label{eq:gauge_transformation}
\end{split}
\end{align}
where $A$ and $B$ are the two sub-lattices of the square lattice. A similar transformation on the creation operators of $\downarrow$ spin particles is performed to maintain the fermionic anti-commutation relations. This gauge transformation gives two bands with opposite curvature at $\textbf{k} =0$ as shown in Fig \ref{fig:band_structure_steps}(b). To make them touch each other, they need to be shifted vertically in opposite directions. This can be achieved by adding a Zeeman field of a specific strength equal to half of the band width, $h=4t$ to the Hamiltonian. Thus, the quadratic part of the microscopic Hamiltonian is
\begin{multline}
H_0  =  \sum_{\langle i,j \rangle} \left( t c_{i\uparrow}^\dagger c_{j\uparrow} -t c_{i\downarrow}^\dagger c_{j\downarrow} + \text{h.c.} \right) \\
-h\sum_{i} \left( c_{i\uparrow}^\dagger c_{i\uparrow} - c_{i\downarrow}^\dagger c_{i\downarrow} \right) 
\nonumber
\end{multline}
where $h=4t$. The dispersions of the bands are $\varepsilon_{\textbf{k}\uparrow} = 2t \cos (k_x) +2t \cos (k_y) -h$ and $\varepsilon_{\textbf{k}\downarrow} = -\varepsilon_{\textbf{k}\uparrow}= -2t \cos (k_x) -2t \cos (k_y) +h$. The spin of the fermions merely acts as a label for the two bands. The results of this paper would remain the same  if a label other than spin was used to label the bands. If we choose $h = 4t$, then it is evident that for small values of $k$, the dispersions become  $\varepsilon_{\textbf{k}\uparrow}  \approx -tk^2$ and $\varepsilon_{\textbf{k}\downarrow} \approx tk^2$. Hence this microscopic model has two bands touching each other isotropically and quadratically as shown in Fig \ref{fig:band_structure_steps}(c). At half filling where the Fermi energy is zero, the lower band of $\uparrow$ spin fermions  is completely filled and the upper band of $\downarrow$ spin fermions is completely empty. Thus, the non-interacting ground state at half-filling has all spins pointing in the same direction, \textit{i.e.} it is ferromagnetic.

The onsite interaction $H_U$ is given by
\begin{equation}
H_U =  U \sum_i n_{i\uparrow} n_{i\downarrow}
\nonumber
\end{equation}
Here, $n_{i\sigma} = c_{i\sigma}^\dagger c_{i\sigma}$ is the number operator on site $i$ with spin $\sigma$. The full Hamiltonian is the sum of the quadratic part and the quartic interactions, $H=H_0+H_U$. The low energy action in imaginary time can be obtained by replacing the annihilation and creation operators by Grassmann fields. This gives the quadratic part  of the low-energy action as
\begin{equation}
S_0 = \int_{|\textbf{k}| < \Lambda,\omega} 
\psi^\dagger(\textbf{k}\omega)
\begin{pmatrix}
-i\omega + tk^2 & 0 \\
0 & -i \omega - t k^2
\end{pmatrix} \psi(\textbf{k}\omega)
\nonumber
\end{equation}
and the quartic part of the action as
\begin{equation} 
S_I = -U \int_{\textbf{k},\omega} \bar{\psi}_\downarrow(4) \bar{\psi}_\uparrow(3) \psi_\downarrow(2) \psi_\uparrow(1)
\nonumber
\end{equation}
where $\psi(\textbf{k}\omega) = \begin{pmatrix}
\psi_{\downarrow} (\textbf{k}\omega) & \psi_{\uparrow} (\textbf{k}\omega)
\end{pmatrix}^T$. This action is the same as that of model B, if we take the two species of fermions of model B as up and down spin fermions, $t\propto 1/m$ and $U \propto -g$.
From the flow Eq. (\ref{eq:g_flow_specific2}), we conclude that repulsive interaction ($U>0$) is  marginally relevant while attractive interaction ($U<0$) is marginally irrelevant in this microscopic model.

Note that the Zeeman field needs to be fine tuned to $4t$ for the bands to touch quadratically. A deviation from this value either creates a gap or leads to a circular Fermi surface where the two bands intersect linearly This Lifshitz transition as the Zeeman field crosses $4t$ is studied in Sec. \ref{sec:another_lt}. Thus the microscopic model lacks the symmetry protection of the quadratic band touching that is present in some other models~\cite{fradkin2009quadratic, uebelacker2011qbcp, pawlak2015qbcp, murray2014qbcp}.

It turns out that it is possible to find the exact ground state of the microscopic  model for the case of repulsive interactions. In fact, the ground state is the same as the non-interacting ground state as discussed below. It is also possible to calculate the exact Green's function of this model which we show in Sec. \ref{subsec:gf}.

\subsection{Ground state}

The ground state of the non-interacting model at half-filling has the lower band completely filled and the upper band completely empty. This non-interacting ground state also minimizes the interaction energy when the interaction is repulsive ($U>0$).
The non-interacting ground state has a single particle on each site, all having the same spin. Since the interaction contains the product of number of up-spins and down-spins at each site, $H_U$ acting on this state gives 0. Thus, the non-interacting ground state is an eigenstate of $H_U$ with eigenvalue 0. Since, $n_{i\uparrow}$ and $n_{i\downarrow}$ are positive-valued operators, $H_U$ which is a sum of product of these positive operators is also a positive-valued operator for $U > 0$. Therefore, all the eigenvalues of $H_U$ are non-negative and 0 is the smallest eigenvalue. This proves that the non-interacting ground state is not only an eigenstate of $H_U$, but also the ground state. Thus, the ground state in the presence of repulsive interactions is the same as the non-interacting ground state, irrespective of the strength of the interaction. Moreover, the ground state energy in the presence of the interaction is the same as that in the absence of interactions. 
\begin{equation}
E_0 = V \int_{BZ} \frac{d\textbf{k}}{(2\pi)^2} \varepsilon_{\textbf{k}\uparrow} = -hN 
\end{equation}
where $V$ is the area of the system and $N$ is the number of particles.

From the RG analysis of Sec. \ref{sec:rg_analysis}, we know that a repulsive interaction is marginally relevant and causes a flow from the Gaussian fixed point to a different (interacting) fixed point where the effective interaction strength is large. However, the calculations done on the above microscopic model indicate that the ground states described by the two fixed points are the same. However, as will be shown below, the excitation spectrum is different at the two fixed points with the interacting system being gapped. 

Note that in obtaining the microscopic Hamiltonian with quadratic band touching, we set the value of the Zeeman field to be $h = 4t$. However, even for field strengths greater than $4t$, the above arguments to determine the ground state hold and it continues to be a ferromagnet.

\subsection{Green's function} \label{subsec:gf}

Now we calculate the Green's function for the system to investigate the excitations about the ground state. In general, it is not possible to determine the Green's function of an interacting system exactly. However, for this particular system, it is possible to calculate it exactly as we demonstrate below. The knowledge of the exact ground state, where every site is occupied by a $\uparrow$ fermion, makes this possible. 

It is most convenient to calculate the Green's function in momentum space and imaginary time.
\begin{widetext}
\begin{equation}
\langle \mathcal{T} c_{\textbf{k}\sigma}^*(\tau) c_{\textbf{k}'\sigma'}(\tau') \rangle = \begin{cases}\expval{\left[  e^{\tau H} c_{\textbf{k}\sigma}^\dagger  e^{- \tau H} \right]\left[ e^{\tau' H} c_{\textbf{k}'\sigma'} e^{- \tau' H}\right]}{GS} & \text{ for } \tau > \tau'\\
-\expval{ \left[ e^{\tau' H} c_{\textbf{k}'\sigma'} e^{- \tau' H} \right]\left[e^{\tau H} c_{\textbf{k}\sigma}^\dagger  e^{- \tau H}\right] }{GS} & \text{ for } \tau < \tau'
\end{cases} 
\nonumber
\end{equation}
\end{widetext}
where $\mathcal{T}$ is the time ordering operator, $c_{\textbf{k}\sigma}^*(\tau) = e^{\tau H} c_{\textbf{k}\sigma}^\dagger  e^{- \tau H}$ and $\ket{GS}$ is the ground state. Let $\Delta \tau \equiv \tau - \tau'$. Now, $e^{-\tau H} \ket{GS} = e^{-\tau E_0} \ket{GS}$. Hence,
\begin{widetext}
\begin{equation}\label{eq:green's_function}
\langle \mathcal{T} c_{\textbf{k}\sigma}^*(\tau) c_{\textbf{k}'\sigma'}(\tau') \rangle = \begin{cases}e^{\Delta \tau E_0} \expval{ c_{\textbf{k}\sigma}^\dagger  e^{- \Delta \tau H}  c_{\textbf{k}'\sigma'} }{GS} & \text{ for } \tau > \tau'\\
-e^{-\Delta \tau E_0} \expval{  c_{\textbf{k}'\sigma'} e^{ \Delta\tau H} c_{\textbf{k}\sigma}^\dagger  }{GS} & \text{ for } \tau < \tau'
\end{cases} 
\end{equation}
\end{widetext}
There are four components of the Green's function corresponding to the different values of $\sigma$ and $\sigma'$. 

\textit{Case 1}: $\sigma = \uparrow$ and $\sigma' = \uparrow$

First consider $\tau>\tau'$. The quantity that is needed to be evaluated is $\expval{ c_{\textbf{k}\uparrow}^\dagger  e^{- \Delta \tau (H_0 + H_U)}  c_{\textbf{k}'\uparrow} }{GS}$.  The state $c_{\textbf{k}'\uparrow}\ket{GS}$ is clearly an eigenstate of $H_0$ with the eigenvalue $E_0 - \varepsilon_{\textbf{k}'\uparrow}$. This state does not contain any particles with $\downarrow$ spin. Hence, it is also an eigenstate of $H_U$ with the eigenvalue 0.
\begin{equation}
(H_0 + H_U) c_{\textbf{k}'\uparrow}\ket{GS} = (E_0 - \varepsilon_{\textbf{k}'\uparrow})  c_{\textbf{k}'\uparrow} \ket{GS} 
\nonumber
\end{equation}
\begin{equation}
\implies e^{-\Delta \tau (H_0 + H_U)} c_{\textbf{k}'\uparrow}\ket{GS} =  e^{-\Delta \tau (E_0 - \varepsilon_{\textbf{k}'\uparrow}) } c_{\textbf{k}'\uparrow} \ket{GS} 
\nonumber
\end{equation}
\begin{equation}\label{eq:case1_condition1}
\implies e^{\Delta \tau E_0} \expval{ c_{\textbf{k}\uparrow}^\dagger  e^{- \Delta \tau H}  c_{\textbf{k}'\uparrow} }{GS} =  e^{\Delta \tau  \varepsilon_{\textbf{k}\uparrow} } \delta_{\textbf{k},\textbf{k}'}
\end{equation}
Now consider $\tau < \tau'$. The quantity that is needed to be evaluated is $\expval{  c_{\textbf{k}'\uparrow} e^{ \Delta\tau H} c_{\textbf{k}\uparrow}^\dagger  }{GS}$. In the ground state all sites are occupied by $\uparrow$ spins and no further $\uparrow$ spin can be accommodated. This implies 
\begin{equation} \label{eq:case1_condition2}
c_{\textbf{k}\uparrow}^\dagger  \ket{GS} = 0
\end{equation}
Using Eq. (\ref{eq:case1_condition1}) and Eq. (\ref{eq:case1_condition2}) in Eq. (\ref{eq:green's_function}) gives 
\begin{equation} \label{eq:case1} 
\langle \mathcal{T} c_{\textbf{k}\uparrow}^*(\tau) c_{\textbf{k}'\uparrow}(\tau') \rangle = \begin{cases} e^{\Delta \tau  \varepsilon_{\textbf{k}\uparrow} } \delta_{\textbf{k},\textbf{k}'} & \text{ for } \tau > \tau'\\
0 & \text{ for } \tau < \tau'
\end{cases} 
\nonumber
\end{equation}

\textit{Case2}: $\sigma = \downarrow$ and $\sigma' = \downarrow$

 Again first consider $\tau>\tau'$. We need to calculate $\expval{ c_{\textbf{k}\downarrow}^\dagger  e^{- \Delta \tau H}  c_{\textbf{k}'\downarrow} }{GS}$. Since the ground state does not contain any $\downarrow$ particles
\begin{equation} \label{eq:case2_condition1}
c_{\textbf{k}'\downarrow} \ket{GS} = 0
\end{equation}

If $\tau<\tau'$, then we need to evaluate $\expval{  c_{\textbf{k}'\downarrow} e^{ \Delta\tau H} c_{\textbf{k}\downarrow}^\dagger  }{GS}$. The state  $c_{\textbf{k}\downarrow}^\dagger  \ket{GS}$ is clearly an eigenstate of $H_0$ with the eigenvalue $E_0+\varepsilon_{\textbf{k}\downarrow}$. Moreover, it is also an eigenstate of $H_U$. 
\begin{equation}
H_U c_{\textbf{k}\downarrow}^\dagger  \ket{GS} = \frac{1}{\sqrt{N}}\sum_j e^{i\textbf{r}_j\cdot \textbf{k}} H_U c_{j\downarrow}^\dagger \ket{GS} 
\nonumber
\end{equation}
Now, $H_U c_{j\downarrow}^\dagger \ket{GS} = Uc_{j\downarrow}^\dagger \ket{GS}$ because the $\downarrow$ fermion in the state $c_{j\downarrow}^\dagger \ket{GS}$ will be repelled by the $\uparrow$ fermion at site $j$. Thus,
\begin{equation}
H_U c_{\textbf{k}\downarrow}^\dagger  \ket{GS} = U c_{\textbf{k}\downarrow}^\dagger  \ket{GS}  
\nonumber
\end{equation}
\begin{equation}
\implies e^{\Delta \tau H} c_{\textbf{k}\downarrow}^\dagger \ket{GS} = e^{\Delta \tau (E_0 + \varepsilon_{\textbf{k}\downarrow} + U)} c_{\textbf{k}\downarrow}^\dagger  \ket{GS} 
\nonumber
\end{equation}
\begin{equation}\label{eq:case2_condition2}
e^{-\Delta \tau E_0} \expval{  c_{\textbf{k}'\downarrow} e^{ \Delta\tau H} c_{\textbf{k}\downarrow}^\dagger  }{GS} =  e^{\Delta \tau (\varepsilon_{\textbf{k}\downarrow} + U)} \delta_{\textbf{k},\textbf{k}'}
\end{equation}
Using Eq. (\ref{eq:case2_condition1}) and Eq. (\ref{eq:case2_condition2}) in Eq. (\ref{eq:green's_function}) gives
\begin{equation} \label{eq:case2} 
\langle \mathcal{T} c_{\textbf{k}\downarrow}^*(\tau) c_{\textbf{k}'\downarrow}(\tau') \rangle = \begin{cases} 0 & \text{ for } \tau > \tau'\\
-e^{\Delta \tau (\varepsilon_{\textbf{k}\downarrow} +U)} \delta_{\textbf{k},\textbf{k}'} & \text{ for } \tau < \tau'
\end{cases}
\end{equation}

\textit{Case3}: $\sigma = \downarrow$ and $\sigma' = \uparrow$

From the conjugate of Eq. (\ref{eq:case2_condition1}),  $\expval{ c_{\textbf{k}\downarrow}^\dagger  e^{- \Delta \tau H}  c_{\textbf{k}'\uparrow} }{GS}=0$. Similarly, from Eq. (\ref{eq:case1_condition2}), $\expval{  c_{\textbf{k}'\uparrow} e^{ \Delta\tau H} c_{\textbf{k}\downarrow}^\dagger  }{GS}=0$. Hence, for this case the Green's function is 0.
\begin{equation} \label{eq:case3} 
\langle \mathcal{T} c_{\textbf{k}\downarrow}^*(\tau) c_{\textbf{k}'\uparrow}(\tau') \rangle = 0
\end{equation}

\textit{Case4}: $\sigma = \uparrow$ and $\sigma' = \downarrow$

Using similar arguments as in case 3, the Green's function for this case is also 0. 
\begin{equation} \label{eq:case4} 
\langle \mathcal{T} c_{\textbf{k}\uparrow}^*(\tau) c_{\textbf{k}'\downarrow}(\tau') \rangle = 0
\end{equation}

Now that the time dependent Green's function has been calculated, we perform Fourier transform to get the following expression for the frequency dependent Green's function. 
\begin{equation}\label{eq:exact_gf_microscopic}
G(\textbf{k},\omega;\textbf{k}',\omega') = 2\pi \delta(\omega - \omega')\delta_{\textbf{k},\textbf{k}'} \begin{pmatrix}
\frac{1}{-i\omega+\varepsilon_{\textbf{k}\downarrow} + U} & 0\\
0 & \frac{1}{-i\omega + \varepsilon_{\textbf{k}\uparrow}}
\end{pmatrix}
\end{equation}
The convention for the basis of the matrix above is as follows: the 11 element corresponds to the $\downarrow\downarrow$ component, the 12 element corresponds to $\downarrow \uparrow$, the 21  element corresponds to $\uparrow \downarrow$ and the 22 element corresponds to $\uparrow \uparrow$.

From the Green's function it can be seen that the dispersion for the $\uparrow$ fermions remains $\varepsilon_{\textbf{k}\uparrow}$. However, the dispersion for the $\downarrow$ particles shifts by a constant amount $U$. The self energy of the $\downarrow$ fermions is a real constant $\Sigma_{\downarrow \downarrow} = U$. The imaginary part of the Green's function gives the spectral function. Here, the spectral function is a delta function indicating that there are well defined excitations in the system and they do not decay. 

We expect that for small $k$, the Green’s function calculated from the field theory, Eq. (\ref{eq:gf_field_theory}) will agree with the one calculated from the microscopic model, Eq. (\ref{eq:exact_gf_microscopic}). Indeed the two agree if we take $U$ to be equal to $-g \Lambda^2 /8 \pi$. The negative sign makes sense because for repulsive interactions, $U > 0$ and $g < 0$.

\subsection{Excitations}

From the form of the Green's function, it is clear that the repulsive interaction changes the dispersion of the $\downarrow$ particles by a constant amount, while the dispersion of $\uparrow$ particles remains unchanged. More importantly, the excitations seem to be stable (infinitely long-lived). We can understand this as follows: At half filling, there is a $\uparrow$ particle on each site. If we remove a $\uparrow$ particle with momentum $\textbf{k}$, the energy required to do so will be  $\varepsilon_{\textbf{k}\uparrow}$ as there are no $\downarrow$ particles to interact with. Hence the dispersion of the $\uparrow$ particles remains unchanged. Now, suppose we add a $\downarrow$ particle to the system at half filling, then the $\uparrow$ particles present on every site act as a uniform background charge. Hence, the $\downarrow$ particle experiences a constant repulsion independent of its position and the dispersion of $\downarrow$ particles shifts by a constant amount $U$.

\section{Another Lifshitz transition} \label{sec:another_lt} 

Till now, we have studied the Lifshitz transition as the carrier density is changed. Another way in which a Lifshitz transition can occur in the microscopic model is by varying the Zeeman field $h$. For $h<4t$, the two bands cross each other and the Fermi energy lies at the intersection of these bands. In this case the Fermi surface is a closed curve. If we assume that $h$ is smaller than $4t$ but close to $4t$, then the Fermi surface is approximately a circle. We will show that in such a situation, the system shows the BCS instability if intraband attractive interactions are present and an instability towards excitonic condensate if interband forward-scattering interactions are present. For $h=4t$, the bands touch quadratically and an attractive interaction is marginally irrelevant while a repulsive interactions is marginally relevant as shown earlier in Sec. \ref{sec:rg_analysis}. For $h>4t$, the bands have a gap and the system is an insulator. Thus the topology of the Fermi surface changes as $h$ crosses $4t$.

In this section, we consider the case $h\lesssim 4t$, where the two bands of model B intersect with each other and the band dispersions are isotropic. Linearizing the bands about the Fermi energy gives the Green's function as
\begin{equation}
    G(\textbf{K},\omega) = 
    \begin{pmatrix}
        \frac{1}{-i\omega + ck} & 0\\
        0 & \frac{1}{-i\omega-ck}
    \end{pmatrix}
\end{equation}
where $c$ is the Fermi velocity and $k$ is measured about the Fermi wave-vector, that is $k = |\textbf{K}|-K_F$.

The general interacting action contains two terms corresponding to the intraband interaction denoted by $S_1$ and $S_2$ and another term corresponding to the interband interaction denoted by $S_{12}$. Note that since the Fermi surface is now a circle, there are coupling \emph{functions} rather than coupling constants.

The three parts of the interacting action are given by
\begin{equation}
    S_1 = \int_{\textbf{k},\omega} g_1 (4321) \bar{\psi}_\downarrow(4) \bar{\psi}_\downarrow(3) \psi_\downarrow(2)\psi_\downarrow(1)
\end{equation}
\begin{equation}
    S_2 = \int_{\textbf{k},\omega} g_2 (4321) \bar{\psi}_\uparrow(4) \bar{\psi}_\uparrow(3) \psi_\uparrow(2)\psi_\uparrow(1)
\end{equation}
\begin{equation}
    S_{12} = \int_{\textbf{k},\omega} g_{12} (4321) \bar{\psi}_\downarrow(4) \bar{\psi}_\uparrow(3) \psi_\downarrow(2)\psi_\uparrow(1)
\end{equation}
The coupling functions $g_1, g_2$ and $g_{12}$ can be either forward-scattering or backward-scattering. The total interaction action is the sum of the three pieces, $S_I = S_1 + S_2 + S_{12}$.
The flow of the couplings is determined by evaluating the following
\begin{equation}
    \langle S_I^2 \rangle = \langle S_1^2 \rangle+ \langle S_2^2 \rangle+2 \langle S_1S_2 \rangle+\langle S_{12}^2 \rangle +2\langle S_1 S_{12}  \rangle + 2\langle S_2S_{12} \rangle
\end{equation}
where it is understood that the terms corresponding to disconnected diagrams are not to be considered.\\ 

\subsection{Intraband couplings}

We now determine the flow of the intraband couplings $g_1$ and $g_2$. The three cross-terms $\langle S_1 S_{12} \rangle$, $\langle S_2 S_{12} \rangle$ and $\langle S_{1}S_2 \rangle$ do not renormalize the intraband couplings $g_1$ and $g_2$, because they involve the off-diagonal components of the Green's function, $G_{\uparrow\downarrow}$ and $G_{\downarrow\uparrow}$ which are zero. It remains to determine the contribution to the flow from the remaining three terms, namely $\langle S_1^2 \rangle$, $\langle S_2^2 \rangle$ and $\langle S_{12}^2 \rangle$.
We consider the cases of forward-scattering couplings and backward-scattering couplings separately. When $g_1$ and $g_2$ are forward-scattering couplings, they will be denoted by $F_1$ and $F_2$. Similarly, when $g_1$ and $g_2$ are backward-scattering couplings, they will be denoted by $V_1$ and $V_2$.

From Shankar's work \cite{shankar1994rg}, we know that the forward-scattering couplings do not get renormalized by the ZS$'$ and the BCS diagrams, for kinematical reasons. More specifically, these diagrams have a  momentum transfer of the order of $K_F$ at the vertices which restricts the integration range of the fast modes to a region of size $d\Lambda^2$.

The calculation of the flow of $F_j$ caused by $\langle S_j^2 \rangle$ would follow exactly like in a Fermi liquid with a single band. From the well known result that the forward-scattering couplings are marginal in a Fermi liquid \cite{shankar1994rg}, it follows that $\langle S_1^2 \rangle$ and $\langle S_2^2 \rangle$ do not renormalize $F_j$. $\langle S_{12}^2 \rangle$ also does not renormalize $F_j$  because the two Green's functions involved in the contribution of the ZS diagram have poles in the same half of the $\omega$ plane. 
Hence, the $F_j$ couplings are marginal and the flow equations are simply
\begin{equation}
    \frac{dF_j}{d\ln{s}} = 0 \;\; \text{ for }j=1,2
\end{equation}

Backward-scattering couplings, $V_1$ and $V_2$  can only get renormalized by BCS diagrams because the ZS and ZS$'$ diagrams have a momentum transfer of order $K_F$ and hence are kinematically suppressed \cite{shankar1994rg}.
The BCS diagram coming from $\langle S_{12}^2 \rangle$ does not renormalize $V_1$ and $V_2$ because it does not correspond to a term of the form $\bar{\psi}_j \bar{\psi}_j \psi_j \psi_j$ for $j \in \{\uparrow,\downarrow\}$.
Thus $V_j$ is renormalized only by $\langle S_j^2 \rangle$ and the calculation proceeds exactly like in a Fermi liquid with a single band. Since we have assumed that $h$ is close to $4t$, the bands dispersions are independent of the direction of the momentum and the system has rotational symmetry. Hence $V_j(4321)$ depends only on the angle between the momenta $\textbf{K}_1$ and $\textbf{K}_3$, that is $V_j(4321) = V_j(\theta_1-\theta_3)$.
If we define $V_{j,l}$ for $j \in \{1,2\}$ and integer values of $l$ as follows
\begin{equation}
    V_{j,l} = \int_0^{2\pi} \frac{d\theta}{2\pi} e^{il\theta} V_j (\theta)
    \nonumber
\end{equation}
then the RG flow equations for $V_{j,l}$ is given by
\begin{equation}
    \frac{dV_{j,l}}{d \ln{s}} = \frac{V_{j,l}^2}{\pi c}
\end{equation}
Negative $V_{j,l}$  corresponds to repulsive interactions and are marginally irrelevant, positive $V_{j,l}$  corresponds to attractive interactions and are marginally relevant leading to an instability towards a superconducting state.

To summarize, the RG flow equations of the intraband couplings are exactly like that of a Fermi liquid.

\subsection{Interband couplings}

\begin{figure}
	\centering
	\includegraphics[width=\columnwidth]{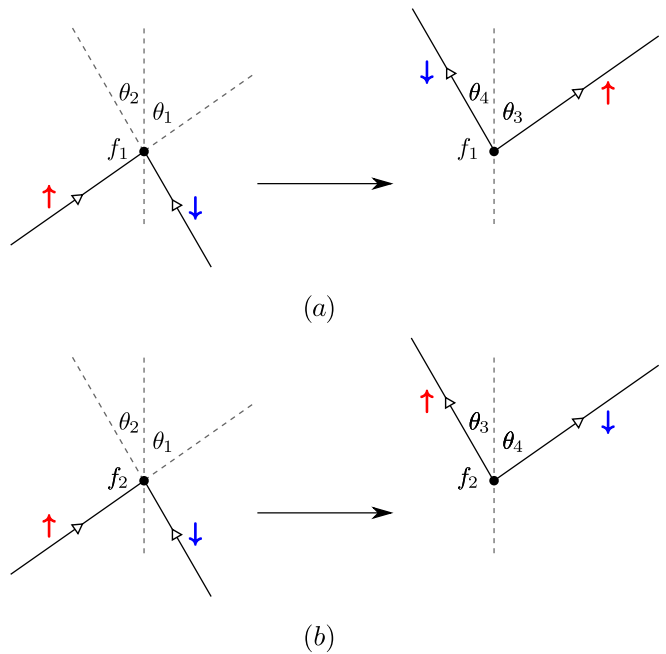}
	\caption{(a) The scattering process for $f_1$. The annihilated $\uparrow$ and created $\uparrow$ particles have the same momentum. (b) The scattering process for $f_2$. The annihilated $\uparrow$ and created $\downarrow$ particles have the same momentum.}
	\label{fig:f1f2}
\end{figure}

\begin{figure}
	\centering
    \bigskip
	\includegraphics[width=0.65\columnwidth]{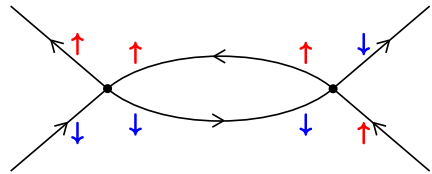}
	\caption{ZS$'$ diagram which gives the flow of interband forward-scattering, $f_2$.}
	\label{fig:zsp2}
\end{figure}

Now we determine the flow of interband coupling function, $g_{12}$. The terms $\langle S_1^2 \rangle$ and $\langle S_2^2 \rangle$ clearly cannot renormalize $g_{12}$ because these terms do not contain both type of fermions. Additionally, $\langle S_1 S_2 \rangle$  does not renormalize $g_{12}$ because it involves the off-diagonal component of the Green's function.  When $g_{12}$ is backward-scattering, we will denote it by $v$. If $g_{12}$(4321) is forward-scattering, then either $\theta_1 = \theta_3$; $\theta_2 = \theta_4$ or $\theta_1 = \theta_4$; $\theta_2 = \theta_3$, where $\theta_i = \tan^{-1}(K_{i,y}/K_{i,x})$. In the former case, we denote $g_{12}$ by $f_1$ and in the latter case, we denote $g_{12}$ by $f_2$. Note that $f_1$ and $f_2$ are independent because $g_{12}(4321) \neq -g_{12}(3421)$ in general. This can be understood by considering the following equality obtained by interchanging the dummy variables $4$ and $3$ followed by anti-commutation of the Grassmann fields.
\begin{align}
   S_{12}&= \int_{\textbf{k},\omega} g_{12} (4321) \bar{\psi}_\downarrow(4) \bar{\psi}_\uparrow(3) \psi_\downarrow(2)\psi_\uparrow(1) \nonumber \\
   & = -\int_{\textbf{k},\omega} g_{12} (3421) \bar{\psi}_\uparrow(4) \bar{\psi}_\downarrow(3) \psi_\downarrow(2)\psi_\uparrow(1) 
   \nonumber
\end{align}
If all the Grassmann fields in the above action had the same subscripts, we would have concluded that $g_{12}(4321) = -g_{12}(3421)$ and $f_1$  would be related to $f_2$ by a minus sign. This is the case for $S_1$ and $S_2$. However, $S_{12}$ involves both the species of Fermions and hence $g_{12}(4321)$ and $g_{12}(3421)$ are independent. This explains why there are two types of interband forward-scattering couplings $f_1$ and $f_2$. The scattering processes corresponding to $f_1$ and $f_2$ are shown in Fig. \ref{fig:f1f2}.

In the case of forward-scattering, it can be shown that diagrams generated by the terms $\langle S_1 S_{12} \rangle$ and $\langle S_{2} S_{12} \rangle$ do not contribute to a flow of $f_1$ and $f_2$  either for kinematical reasons or because the  $\omega$ integral vanishes. Now it remains to analyze the contribution to the flow from $\langle S_{12}^2 \rangle$.

For the flow of $f_1$, the ZS$'$ and BCS diagrams coming from $\langle S_{12}^2 \rangle$ are kinematically suppressed and the ZS diagram involves the off-diagonal components of the Green's function. Thus $f_1$ is marginal and its RG flow equation is
\begin{equation}
\frac{d f_1}{d \ln s} = 0
\end{equation}

For the flow of $f_2$, the ZS and BCS diagrams are kinematically suppressed and the flow is given by the ZS$'$ diagram coming form $\langle S_{12}^2 \rangle$ which is shown in Fig. \ref{fig:zsp2}.
\begin{widetext}
    \begin{align}
        df_2(4321) = \int_{d\Lambda} \frac{dK}{2\pi} \int_{0}^{2\pi} \frac{d\theta}{2\pi}f_2(4,K,K,1) f_2(K,3,2,K) \int_{-\infty}^{+\infty} \frac{d\omega}{2\pi} G_{\downarrow \downarrow}(K,\omega)G_{\uparrow \uparrow}(K,\omega)
    \end{align}
    \nonumber
\end{widetext}
The poles of $G_{\uparrow\uparrow}(K,\omega)$ and $G_{\downarrow\downarrow}(K,\omega)$ lie in the opposite halves of the $\omega$ plane and hence the $\omega$ integral is non-zero.  Because of the rotational symmetry for $h$ close to $4t$, $f_2$ must depend only on the difference $\theta_1 - \theta_2$. If we define $f_{2,l}$ as follows
\begin{equation}
f_{2,l} = \int_0^{2\pi} \frac{d\theta}{2 \pi} e^{il\theta} f_{2,l}(\theta)
\nonumber
\end{equation}
then the flow equation for $f_{2,l}$ is
\begin{equation}
\label{eq:f2l_flow}
\frac{d f_{2,l}}{d\ln s} = -\frac{f_{2,l}^2}{2 \pi c} 
\end{equation}We see that negative $f_{2,l}$ is marginally relevant and positive $f_{2,l}$ is marginally irrelevant. The instability that the negative $f_{2,l}$ correspond to is the excitonic condensate instability. This instability is discussed in the next section.

For backward-scattering, the ZS and ZS$'$ diagrams are kinematically suppressed and only BCS diagrams can give a flow. The BCS diagrams from $\langle S_1 S_{12} \rangle$ and $\langle S_2 S_{12}\rangle$ do not cause a flow of $v$ because they generate terms of the form $\int \bar{\psi}_\downarrow \bar{\psi}_\downarrow \psi_\uparrow \psi_{\uparrow}$ and $\int \bar{\psi}_\uparrow \bar{\psi}_\uparrow \psi_\downarrow \psi_{\downarrow}$, which are different from the form of $S_{12}$. We saw that for the case of intraband couplings, $\langle S_j^2\rangle$ generated a flow of $V_j$. Naively one would expect that $\langle S_{12}^2 \rangle$ will cause a flow of $v$. However, this turns out to be incorrect as shown below. The change in $v$ is given by the contribution of the BCS diagram coming from $\langle S_{12}^2 \rangle$, which is

\begin{widetext}
    \begin{align}
        dv(4321) = \int_{d\Lambda} \frac{dK}{2\pi} \int_{0}^{2\pi} \frac{d\theta}{2\pi}v(-K,K,2,1) v(4,3,-K,K) \int_{-\infty}^{+\infty} \frac{d\omega}{2\pi} G_{\downarrow\downarrow}(-K,-\omega)G_{\uparrow\uparrow}(K,\omega)
    \end{align}
\end{widetext}
The $\omega$ integral is as follows
\begin{multline}
    \label{eq:v12_bcs}
    \int_{-\infty}^{+\infty} \frac{d\omega}{2\pi} G_{\downarrow\downarrow}(-K,-\omega)G_{\uparrow\uparrow}(K,\omega)=\\
    \int_{-\infty}^{+\infty} \frac{d\omega}{2\pi} \frac{1}{(i\omega+ck)}\frac{1}{(-i\omega-ck)} =0
\end{multline}
This integral is 0 because both the poles of the integrand lie in the same half of the $\omega$ plane. Hence the term $\langle S_{12}^2 \rangle$ also does not renormalize $g_{12}$ and $v$ is marginal.
\begin{equation}
\frac{d v}{d \ln s} = 0
\end{equation}
Note that for the above $\omega$ integral to be 0, it is important that the signs of the Fermi velocities in the two Green's functions $G_{\uparrow\uparrow}$ and $G_{\downarrow\downarrow}$ be opposite. 

There is an intuitive way to understand the reason behind the absence of instability in the interband backward-scattering channel.
In a system with a single band, the instability is caused because it is energetically favourable to form a Cooper pair.
However, for our system it is not possible to form a Cooper pair in which the two particles are from different bands. To see this, note that $\downarrow$ band is empty for $|\textbf{K}|>K_F $ while band $\uparrow$ is empty for $|\textbf{K}|< K_F$. If we attempt to create a Cooper pair in which one of the particles is from band $\downarrow$ and  has momentum $\textbf{K}$, whose magnitude is greater than $K_F$, then its partner must have momentum $-\textbf{K}$ and should be from band $\uparrow$. However, the state with momentum $-\textbf{K}$ in band $\uparrow$ is already occupied (because $|-\textbf{K}|> K_F$). Hence a Cooper pair in which one of the particle is from band $\downarrow$ and the other is from band $\uparrow$ cannot be formed and there is no instability. This argument holds good as long as the signs of the Fermi velocities of the two bands are different which is consistent with the requirement for the integral in Eq. (\ref{eq:v12_bcs}) to be zero. 

\subsection{Exciton condensate instability}

The equation (\ref{eq:f2l_flow}) shows that the system has an instability because of $f_2$ coupling. We will show that the instability is towards the formation of an excitonic condensate. Consider the following interaction.
\begin{equation}
    \label{eq:exciton_interaction}
    H_I = \frac{U}{V} \sum_{\textbf{k},\textbf{k}'} c_{\textbf{k}\uparrow}^\dagger c_{\textbf{k}' \downarrow}^\dagger c_{\textbf{k}\downarrow} c_{\textbf{k}' \uparrow}
    \nonumber
\end{equation}
where $V$ is the area of the system.  $\textbf{k}$ and $\textbf{k}'$ above are restricted so that $|\varepsilon_{\textbf{k}}|<c\Lambda$ and $|\varepsilon_{\textbf{k}'}|<c\Lambda$. This interaction corresponds to $f_2$ because the momentum of the annihilated $\uparrow$ particle is the same as the momentum of the created $\downarrow$ particle and the momentum of the annihilated $\downarrow$ particle is the same as the momentum of the created $\uparrow$ (See Fig. \ref{fig:f1f2} (b)).
It is instructive to write the Hamiltonian in terms of particle operator, $c_{\textbf{k}\uparrow}$ and hole operators $a_{\textbf{k}\downarrow} = c_{-\textbf{k}\downarrow}^\dagger$ to understand the instability that the above perturbation leads to.
The non-interacting part of the Hamiltonian in terms of these operators is
\begin{align}
    H_0 & = \sum_\textbf{k} \varepsilon_{\textbf{k}} c_{\textbf{k}\uparrow}^\dagger c_{\textbf{k}\uparrow} - \varepsilon_{\textbf{k}} c_{\textbf{k}\downarrow}^\dagger c_{\textbf{k}\downarrow} \nonumber\\
    & = \sum_\textbf{k} \varepsilon_{\textbf{k}} c_{\textbf{k}\uparrow}^\dagger c_{\textbf{k}\uparrow} + \varepsilon_{\textbf{k}} a_{\textbf{k}\downarrow}^\dagger a_{\textbf{k}\downarrow} - \sum_{\textbf{k}} \varepsilon_{\textbf{k}}
\end{align}
The interacting part of the Hamiltonian can be written as
\begin{equation}
    H_I = -\frac{U}{V} \sum_{\textbf{k},\textbf{k}'}c_{\textbf{k}\uparrow}^\dagger a_{-\textbf{k} \downarrow}^\dagger a_{-\textbf{k}'\downarrow} c_{\textbf{k}' \uparrow} = -\frac{U}{V} A^\dagger A
    \nonumber
\end{equation}
where $A^\dagger = \sum_{\textbf{k}} c_{\textbf{k}\uparrow}^\dagger a_{-\textbf{k}\downarrow}^\dagger$ is the creation operator of excitons and $A = \sum_{\textbf{k}}a_{-\textbf{k}\downarrow} c_{\textbf{k}\uparrow} $ is the annihilation operator of excitons.
This Hamiltonian is exactly like the BCS Hamiltonian except it contains hole creation and annihilation operators.
If we define $\Delta = -U \langle A \rangle/V$, then under the mean-field approximation, $A^\dagger A \approx A^\dagger \langle A \rangle + \langle A^\dagger \rangle A - \langle A^\dagger \rangle \langle A \rangle$, the Hamiltonian (up to a constant) becomes
\begin{equation}
    H = \psi_{\textbf{k}}^\dagger 
    \begin{pmatrix}
        \varepsilon_{\textbf{k}} & \Delta\\
        \Delta^* & -\varepsilon_{\textbf{k}}
    \end{pmatrix}
    \psi_{\textbf{k}}
    \nonumber
\end{equation}
where $\psi_{\textbf{k}}^\dagger = \begin{pmatrix} c_{\textbf{k}\uparrow}^\dagger & a_{-\textbf{k}\downarrow} \end{pmatrix} $ is Nambu spinor. Evaluating the expectation value of $A$ in the ground state of the above Hamiltonian and substituting it in the definition of $\Delta$, which ensures self-consistency gives the famous BCS gap equation \cite{coleman2015}.
\begin{equation}
\Delta = \frac{U}{2V} \sum_{\textbf{k}} \frac{\Delta}{\sqrt{\varepsilon_\textbf{k}^2+|\Delta|^2}}
\end{equation}
This equation has a non-trivial solution
\begin{equation}
    \Delta = 2 c\Lambda e^{-\frac{1}{U N(0)}}
\end{equation}
where $N(0)$ is the density of states at the Fermi level. This shows that the $f_2$ interaction leads an instability towards an excitonic condensate \cite{keldysh1968exciton}.

\section{Discussion} \label{sec:discussion}

We have studied the effect of short-ranged interactions on two low-energy models having single quadratic band touching point using the Renormalization Group. We have shown that the system has only one coupling constant when the Fermi energy is at the band touching point. The flow equations for this coupling constant at the one loop level for both models are identical. A repulsive interaction is marginally relevant while an attractive interaction is marginally irrelevant. We have also calculated the Green's function to one-loop order for both models.

We have studied a microscopic model which in the low energy limit yields the model B in our analysis. When the Fermi energy is exactly at the band touching point, this model turns out to be exactly solvable. Moreover, we are able to calculate the Green's function for the exactly solvable model. The study of the microscopic model also sheds light on the result that repulsive interactions are marginally relevant obtained from the field theoretic calculation. It shows that the ground states of the non-interacting model and the one with repulsive interactions are identical. This implies that the interacting fixed point that the flow is directed towards has the same ground state as the Gaussian fixed point. Thus, there is presumably no change of the ground state along the flow even though the coupling changes. However, the spectrum of excitations is different at the two fixed points with there being a gap at the interacting fixed point but none at the non-interacting (Gaussian) one. 

In the case of attractive interactions, if we flip a fraction $n$ of the spins in the ground state, then the excitation energy required is proportional to $tn^2$. The energy lowered by the attractive interaction is proportional to $|U|n(1-n)$. For the flipping of spins to be energetically favourable, we require $tn^2<|U|n(1-n)$, that is $n<\frac{|U|/t}{1+|U|/t}$. Hence, if $|U|/t$ is infinitesimal, then only an infinitesimal fraction of the particles will have $\downarrow$ spin in the interacting ground state. Thus, the interacting ground state is the same as the non-interacting ground state, at least in the limit of weak interactions. This is consistent with the attractive interaction being marginally irrelevant.

The exact Green's function of the microscopic model shows that the excitations even in the presence of interactions are infinitely long-lived. Further, the Green's function matches that obtained from the field theoretic calculation of the continuum model to one loop, which also seems to have no imaginary part of the self-energy. It is possible that this is a feature of the field theory only to one loop and higher loop corrections will introduce a finite lifetime. This would also signal a point of departure of the continuum model from the microscopic one. 

We have studied another Lifshitz transition in the microscopic model which occurs as the Zeeman field is varied. When the bands intersect each other, we find that attractive intraband backward-scattering couplings, $V_1$ and $V_2$  lead to the BCS instability and the  interband forward-scattering coupling $f_2$ leads to an instability towards an excitonic condensate. The interband backward-scattering coupling $v$, and the intraband forward-scattering couplings, $F_1$ and $F_2$ are found to be marginal.

We have thus identified Lifshitz transitions at the band touching point. The effect of interactions is different at this point as compared to when the Fermi energy is within the bands, and the system is a Fermi liquid. 

\section{Acknowledgments}
We thank Vijay Shenoy for insightful discussions and especially for suggesting the calculation of Sec.~\ref{sec:another_lt}.

\bibliographystyle{apsrevnourl}

\bibliography{references}

\end{document}